\documentclass[superscriptaddress,twocolumn,prl,amsmath,amssymb]{revtex4-2}

\usepackage{amsthm}
\usepackage{graphicx}
\usepackage{color}
\usepackage{dcolumn}
\usepackage{bm}
\usepackage{braket}
\usepackage{float}
\usepackage[table]{xcolor}
\usepackage{enumerate}

\newtheorem{result}{Result}

\newcommand{\cF}{\ensuremath{\mathcal{F}}}
\newcommand{\cG}{\ensuremath{\mathcal{G}}}
\newcommand{\cH}{\ensuremath{\mathcal{H}}}

\newcommand{\cV}{\ensuremath{\mathcal{V}}}
\newcommand{\cK}{\ensuremath{\mathcal{K}}}

\newcommand{\cX}{\ensuremath{\mathcal{X}}}
\newcommand{\cY}{\ensuremath{\mathcal{Y}}}

\newcommand{\cW}{\ensuremath{\mathcal{W}}}

\newcommand{\cU}{\ensuremath{\mathcal{U}}}

\newcommand{\bx}{\ensuremath{\boldsymbol{x}}}
\newcommand{\by}{\ensuremath{\boldsymbol{y}}}

\newcommand{\bap}{\ensuremath{\boldsymbol{\alpha}}}

\newcommand{\ba}{\ensuremath{\boldsymbol{a}}}

\newcommand{\bbN}{\ensuremath{\mathbb{N}}}
\newcommand{\bbR}{\ensuremath{\mathbb{R}}}
\newcommand{\bbQ}{\ensuremath{\mathbb{Q}}}

\DeclareMathOperator{\tr}{\textup{Tr}}

\DeclareMathAlphabet{\mymathbb}{U}{BOONDOX-ds}{m}{n}

\usepackage{hyperref}
\hypersetup{
    colorlinks = true,
    linkcolor = {blue},
    citecolor = {blue},
    urlcolor = {blue},
}

\begin{document}
\title{Universal Approximation Property of Quantum Machine Learning Models\\in Quantum-Enhanced Feature Spaces}

\makeatletter
\def\@fnsymbol#1{\ensuremath{\ifcase#1\or \dagger\or * \or \ddagger\or
   \mathsection\or \mathparagraph\or \|\or **\or \dagger\dagger
   \or \ddagger\ddagger \else\@ctrerr\fi}}
   
\def\frontmatter@thefootnote{%
 \altaffilletter@sw{\@fnsymbol}{\@fnsymbol}{\csname c@\@mpfn\endcsname}%
}%
\makeatother

\author{Takahiro Goto}
\email{goto.takahiro.2020@gmail.com}
\affiliation{
    Reservoir Computing Seminar Group, 
    Nagase Hongo Building F8, 5-24-5, Hongo, Bunkyo-ku, Tokyo 113-0033, Japan
}

\author{Quoc Hoan Tran}
\email{tran\_qh@ai.u-tokyo.ac.jp (Corresponding author)}

\author{Kohei Nakajima}
\email{k\_nakajima@mech.t.u-tokyo.ac.jp}
\affiliation{
    Reservoir Computing Seminar Group, 
    Nagase Hongo Building F8, 5-24-5, Hongo, Bunkyo-ku, Tokyo 113-0033, Japan
}
\affiliation{
	Graduate School of Information Science and Technology, The University of Tokyo, Tokyo 113-8656, Japan
}

\date{\today}

\begin{abstract}
Encoding classical data into quantum states is considered a quantum feature map to map classical data into a quantum Hilbert space. 
This feature map provides opportunities to incorporate quantum advantages into machine learning algorithms
to be performed on near-term intermediate-scale quantum computers. 
The crucial idea is using the quantum Hilbert space as a quantum-enhanced feature space in machine learning models.
While the quantum feature map has demonstrated its capability when combined with linear classification models in some specific applications, its expressive power from the theoretical perspective remains unknown. 
We prove that the machine learning models induced from the quantum-enhanced feature space are universal approximators of continuous functions under typical quantum feature maps.
We also study the capability of quantum feature maps in the classification of disjoint regions. 
Our work enables an important theoretical analysis to ensure that machine learning algorithms based on quantum feature maps can handle a broad class of machine learning tasks.
In light of this, one can design a quantum machine learning model with more powerful expressivity.
\end{abstract}

\pacs{Valid PACS appear here}

\maketitle

The rapidly increasing volume and complexity of data have led to the notable progress of machine learning (ML) techniques to build sophisticated models to find patterns in data. The main interest lies in the ability to recognize the patterns these techniques can produce. 
If a physical computation model can produce atypical patterns that cannot be generated by a classical computer, it may reveal patterns that are difficult to recognize in the classical regime~\cite{biamonte:2017:QML}. This expectation has led to the advent of quantum machine learning (QML), a field that takes advantage of quantum effects to surpass the classical ML techniques. QML is currently benefiting from the arrival of noisy intermediate-scale quantum devices that may include a few tens to hundreds of qubits with no error correction capability~\cite{preskill:2018:NISQ,torlai:2020:NISQ}.
Such devices have ushered in the era of hybrid quantum-classical algorithms~\cite{peruzzo:2014:VQE,farhi:2014:QAOA,fujii:2017:qrc,mitarai:2018:circuit,halvlicek:2019:supervised,schuld:2019:feature}.

\begin{figure}
		\includegraphics[width=9cm]{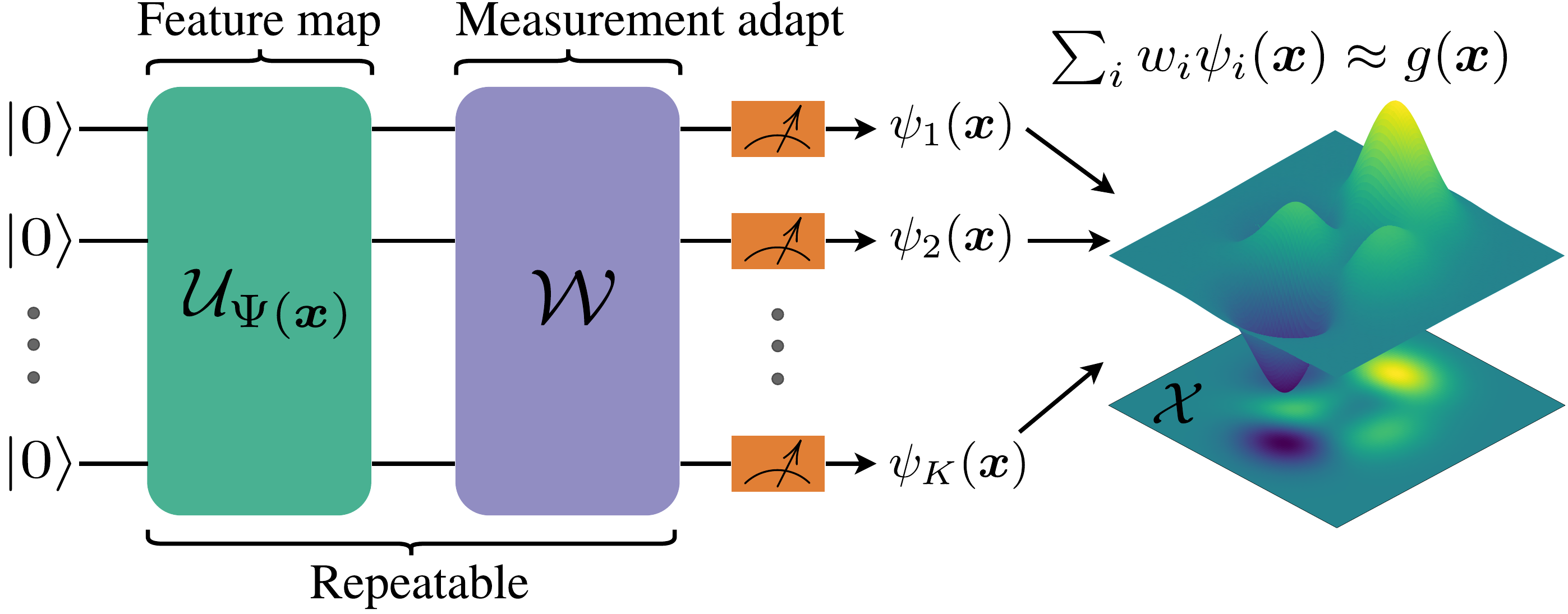}
		\protect\caption{A quantum feature framework consists of a feature map circuit $\cU_{\Psi(\bx)}$ that realizes $\Psi(\bx)$ to map the classical data $\bx\in\cX$ to a quantum state in the Hilbert space and a quantum circuit $\cW$ to adapt the measurement basis. 
		The combination of $\cU_{\Psi(\bx)}$ and $\cW$ can be repeated as a sequence with different parameters.
		This framework has the universal approximation property if the linear combining of measurement results can approximate any continuous function $g:\cX\to\bbR$.
		\label{fig:qft}}
\end{figure}

Because a quantum computer can efficiently access and manipulate quantum states, the quantum Hilbert space can be used as a quantum-enhanced feature space for classical data.
The motivation is that quantum systems can explore a larger class of features than can classical systems.
The input data is encoded in a quantum state via a quantum feature map, a nonlinear feature map that maps data to the quantum Hilbert space (Fig.~\ref{fig:qft}).
The quantum feature map is first proposed and implemented as a fixed quantum circuit, followed by a variational circuit that adapts the measurement basis with trainable parameters~\cite{halvlicek:2019:supervised,schuld:2019:feature}. 
Such QML models can be rephrased as quantum kernel methods induced from feature maps~\cite{park:2020:quantumkernel,blank:2020:taylor,larose:2020:robust,lloyd:2020:quantum,schuld:2021:supervised}. 
Quantum feature maps underscore the QML advantage; there may be a provable exponential speed-up due to the classical intractability of generating correlations for a particular learning problem.
For example, under the widely known hardness assumption of the discrete logarithm problem, the first probable exponential QML advantage was demonstrated via the estimation of a support vector machine kernel matrix on a fault-tolerant quantum computer~\cite{liu:2020:rigorous}.
Furthermore, one can construct engineered data sets to demonstrate the most significant separation between quantum and classical models from a learning-theoretic sense to yield the quantum advantage in ML problems~\cite{huang:2021:power}.
Still, little is known about the relation between the classical intractability of quantum feature maps and the generalization learning performance.

An interesting research question is whether a QML model based on a quantum feature map can obtain expressivity that is as powerful as, or is more powerful than, classical ML schemes.
The answer can determine whether QML models can handle a broad class of ML tasks in general.
This can be investigated from the perspective of the universal approximation property (UAP) and the classification capability, which have been extensively explored in feedforward classical neural networks~\cite{huang:2000:single,huang:2006:universal,huang:2007:convex}.
Here, UAP refers to the ability to approximate any continuous function~\cite{cybenko:1989:sigmoid,hornik:1991:apprx}. The classification capability implies that the function constructed from quantum feature maps can form disjoint decision regions~\cite{huang:1998:neural}.
Quantum neural networks, which employ qubits as quantum perceptrons with nonlinear excitation responses~\cite{torrontegui:2019:QUAP},
can be emulated on a photonic quantum computer to obtain UAP~\cite{killoran:2019:cvmodel}.
It is conjectured that under a special kind of classical data pre-processing, sequentially repeated quantum feature maps can become universal function approximators~\cite{perez:2020:reupload}.
In Ref.~\cite{schuld:2020:encoding}, the expressivity of a quantum model with a variational circuit is characterized in terms of a partial Fourier series in the data.
However, the study of UAP and classification capability of QML models with quantum feature maps still remains challenging.

In this Letter, we formulate the universal approximation problem of QML models in terms of quantum feature maps. 
We present a provable UAP and classification capability in two typical scenarios when setting the quantum feature map. 
In the first scenario, which is defined as the \textit{parallel scenario}, the quantum feature map is a tensor product of multiple quantum circuits; each circuit acts on a subsystem, and the number of qubits can be set freely.
In the second scenario, which is defined as the \textit{sequential scenario}, the quantum feature map is the repetition of a simple fixed quantum circuit, and the number of qubits is fixed.
We obtain the UAP in the first scenario and prove the UAP for the second in single-qubit circuits of the finite input space. 
Both scenarios have been mentioned in prior proposals via short circuit sequences in realistic near-term settings~\cite{halvlicek:2019:supervised,schuld:2020:centric,schuld:2020:encoding}.
We therefore focus on the extent to which these abstract setups can influence the approximating power of QML models in future implementations with wider and deeper quantum circuits.

\textit{Quantum feature maps.---} We will now define the quantum feature map mentioned in Refs.~\cite{halvlicek:2019:supervised,schuld:2019:feature}. Let $\cH$ be a Hilbert space and $\cX \subset \bbR^d$ be an input set.
The quantum feature map $\Psi: \cX \to \cH$ is a procedure of input encoding that encodes some input $\bx\in \cX$ into a quantum feature state $\ket{\Psi(\bx)} \in \cH$.
This mapping action is equivalent to applying the quantum circuit $\cV(\bx)=\cU_{\Psi(\bx)}$ to the initial state $\ket{0}^{\otimes N}$, where $N$ is the number of qubits.
A quantum classifier can be constructed from the quantum feature map using two approaches,  the variational circuit approach and the kernel-induced approach.
In the variational circuit approach, a short-depth quantum circuit $\cW$ is applied to the quantum feature state to adapt the measurement basis~\cite{halvlicek:2019:supervised,schuld:2019:feature} (Fig.~\ref{fig:qft}).
The parameters of circuit $\cW$ are optimized during the training and
the quantum measurement is performed to obtain a complex nonlinear output.
This output can be represented as a linear combination of exponentially many nonlinear functions.
In the kernel-induced approach, the quantum computer estimates the inner product between quantum feature states giving rise to a kernel $\kappa(\bx, \bx^{\prime})=\braket{\Psi(\bx)|\Psi(\bx^\prime)}=\bra{0\ldots0}\cV^{\dagger}(\bx)\cV(\bx^{\prime})\ket{0\ldots0}$ 
to feed into classical kernel methods~\cite{schuld:2019:feature}.

\textit{Quantum feature framework.---} We unify the two above approaches into a quantum feature framework combining  quantum feature maps with an appropriate possible set of observables.
We introduce observables $O_1, O_2, \ldots, O_K$, which are Hermitian operators  applied to the state $\ket{\Psi(\bx)}$.
If we measure $O_i$, we can obtain the expectation value of this observable and consider it as the basis function $\psi_i(\bx): \cX \to \bbR$,
defined as
\begin{align}\label{eqn:basic}
    \psi_i(\bx) = \braket{\Psi(\bx)|O_i|\Psi(\bx)} = \tr[O_i \ket{\Psi(\bx)}\bra{\Psi(\bx)}].
\end{align}
If these basis functions have nonlinearity properties with sufficiently high dimension, we can solve a complex task by the linear regression on the output function $f:\cX \to \bbR$, which is the linear combination of the basis functions $\psi_i(x)$ with the weights $w_i \in \bbR$ $ (i=1,\ldots,K)$~\cite{fujii:2020:quantum}
\begin{align}\label{eqn:linear:qelm}
f(\bx) = \sum_{i=1}^Kw_i\psi_i(\bx).
\end{align}
The observables $\{O_i\}$ should be chosen for easy physical implementation but can produce nonlinearity with sufficient high-dimensional basis functions
\footnote{This scheme is analogous with the classical extreme learning machine (ELM) framework~\cite{huang:2000:single,huang:2006:elm}.
In the ELM, the input data $\boldsymbol{x}$ is fed into a single- or multi-layer perceptron where all weights between layers are fixed. The states of hidden nodes at some layers are regarded as basis functions that play a similar role as $\psi_i(\boldsymbol{x})$.}.

\textit{Universal approximation property and classification capability.---}
A quantum feature framework $\cF$ based on a set of quantum feature maps and a set of observables on the Hilbert space is defined as the collection of function $f: \cX \to \bbR$, where each $f$ has the form in Eq.~\eqref{eqn:linear:qelm}. 
We define the UAP and classification capability of $\cF$.
Let $\cG$ be a space of continuous functions $g:\cX \to \bbR$. The framework $\cF$ has the UAP with respect to $\cG$ and a norm $\Vert \cdot \Vert$ if given any function $g\in\cG$; then for any $\varepsilon > 0$ there exists $f \in \cF$ such that $\Vert f - g \Vert < \varepsilon$. This $f$ is called an approximator of $g$ with $\varepsilon$-error.
Furthermore, $\cF$ has the classification capability if for arbitrary disjoint regions (i.e., closed sets) $\cK_1, \cK_2,\ldots,\cK_m$ in $\cX$, there exists $f\in \cF$ such that $f$ can separate these regions~\cite{huang:2000:single}. 
We investigate the UAP and the classification capability in two typical scenarios in setting the quantum feature map. We assume that $\cX$ is a compact set.
For the sake of readability, we present some definitions for notations used in this study.
A supremum norm of a function $h: \cX \to \bbR$ is defined as $\Vert h \Vert_{\infty} = \sup_{\bx \in \cX} |h(\bx)|$. 
Let $L^2(\cX)$ be a space of functions $h:\cX \to \bbR$ that is square integrable, that is, $\int_{\cX}|h(\bx)|^2d\bx < \infty$.
The norm of function $h$ in $L^2(\cX)$ space is defined as $\Vert h \Vert_{L^2(\cX)}=\left[\int_{\cX}|h(\bx)|^2d\bx \right]^{1/2}$.

\begin{figure}
		\includegraphics[width=9cm]{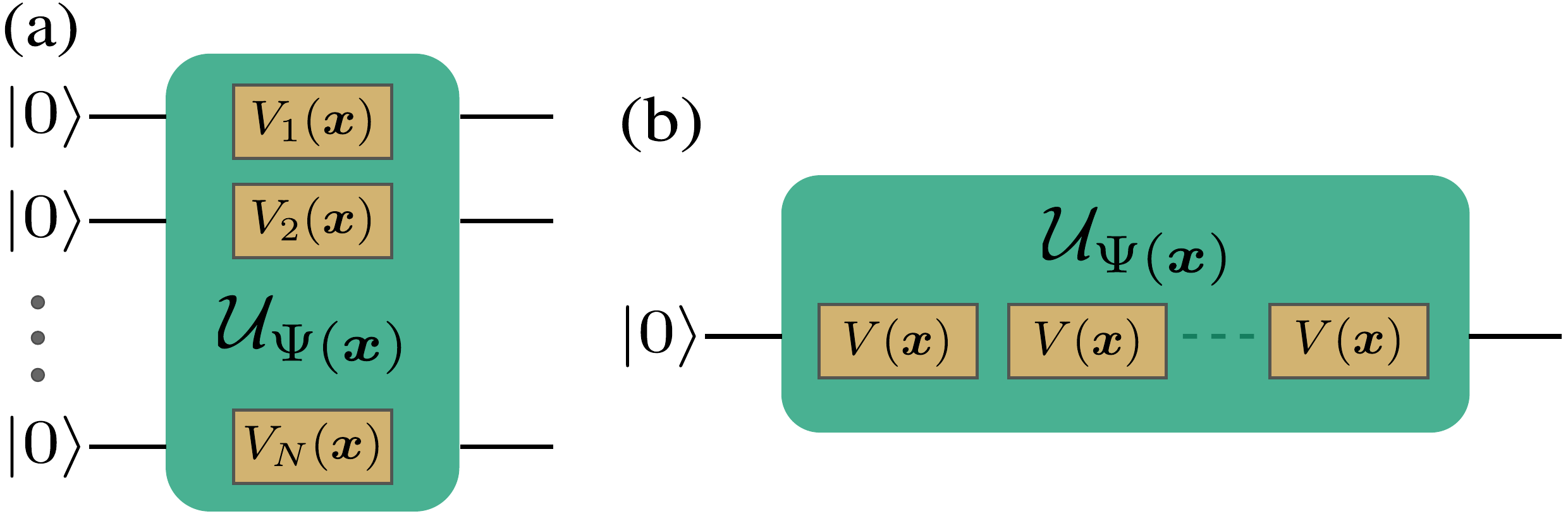}
		\protect\caption{The quantum circuit $U_{\Psi(\bx)}$ for a quantum feature map. (a) The circuit is the tensor product of multiple circuits, where each circuit $V_i(\bx)$ acts on a subsystem. (b) The circuit is the repetition of a simple circuit $V(\bx)$ (for example, a single Pauli-Y rotation) acting on the same qubits.
		\label{fig:circuit}}
\end{figure}

\textit{Parallel scenario.---} 
We examine the first scenario where the quantum feature map is a tensor product of multiple quantum circuits acting on subsystems where the number of qubits can be set freely~[Fig.~\ref{fig:circuit}(a)].
We consider a typical feature map $\Psi_{N}^{\cV}$ represented by 
the following circuit applied to $\ket{0}^{\otimes N}$
\begin{align}\label{eqn:feature:v}
    \cV_N(\bx) = V_1(\bx) \otimes V_2(\bx) \otimes \ldots \otimes V_N(\bx),
\end{align}
where $V_j(\bx)$ is a single-qubit Pauli rotation, for example, $Y$-basis rotation $e^{-i\theta_j(\bx)Y}$ applied to the $j$th qubit with the function $\theta_j:\cX\to\bbR$.
Here, 
$
I = \left[\begin{array}{cc}1 & 0\\0 & 1\end{array}\right]$,
$
X = \left[\begin{array}{cc}0 & 1\\1 & 0\end{array}\right]$,
$
Y = \left[\begin{array}{cc}0 & -i\\i & 0\end{array}\right]$,
and
$
Z = \left[\begin{array}{cc}1 & 0\\0 & -1\end{array}\right]$ 
are the Pauli matrices.
We show that the UAP can be obtained via the nonlinearity of the basis functions.
This nonlinearity can be introduced by an appropriate selection of observables or by a classical pre-processing, such as using a nonlinear pre-transformation for the input.

To begin, we propose a popular setting of $\theta_j(\bx)$ and observables to produce the nonlinearity in the quantum feature framework.
Because $\cX$ is a compact subset of $\bbR^d$, without a loss of generality, we assume that $\cX = [0, 1]^d$.
Given the input data $\bx=(x_1,\ldots,x_d)\in\cX$ and $N\geq d$, we consider the circuits in Eq.~\eqref{eqn:feature:v} with $V_j(\bx) = e^{-i \textup{arccos}(\sqrt{x_k})Y}$,  where $1 \leq k \leq d$ and $k \equiv j (\textup{mod } d)$, $(1\leq j \leq N)$.
The observables are
$
    O_{\bap} = Z^{\alpha_1}\otimes Z^{\alpha_2}\otimes \ldots \otimes Z^{\alpha_N},
$
where $\bap = (\alpha_1, \alpha_2, \ldots, \alpha_N) \in \{0, 1\}^N$.
The basis functions are calculated as
\begin{align}\label{eqn:basis:rs1}
\psi_{\bap}(\bx) 
    &= \bra{0}^{\otimes N} \cV_N^{\dagger}(\bx) O_{\bap} \cV_N(\bx) \ket{0}^{\otimes N}.
\end{align}
From $\{\psi_{\bap}\}$, we can construct any polynomial function on $\cX$~\cite{supp}.
Due to a special case of Stone--Weierstrass theorem~\cite{yoshida:1980:functional}, any continuous function on $\cX$ can be approximated by polynomial functions with arbitrary precision in terms of the supremum norm. Therefore, we obtain the following UAP (see proof in \cite{supp}).

\begin{result}[UAP in the parallel scenario]\label{theo:scene2:uap}
For any continuous function $g: \cX \to \bbR$; 
then for any $\varepsilon>0$, there exist $N$ and a collection of output weights $w_{\bap}$ and observables $O_{\bap} = Z^{\alpha_1}\otimes Z^{\alpha_2}\otimes \ldots \otimes Z^{\alpha_N}$, where $\bap = (\alpha_1, \alpha_2, \ldots, \alpha_N) \in \{0, 1\}^N$ such that
  $
    | \sum_{\bap} w_{\bap} \psi_{\bap}(\bx)  - g(\bx) | < \varepsilon
  $
  for all $\bx$ in $\cX$.
  Here, the basis function $\psi_{\bap}(\bx)$ is defined as that in Eq.~\eqref{eqn:basis:rs1}.
\end{result}

Result~\ref{theo:scene2:uap} implies that the induced quantum feature framework has the UAP with respect to the supremum norm.
Furthermore, we prove the classification capability of this framework.
We consider $m$ disjoint regions $\cK_1, \cK_2,\ldots,\cK_m$ in $\cX$ and their corresponding $m$ distinct real values as labels $c_1, c_2,\ldots,c_m$.
According to lemma 2.1 in Ref.~\cite{huang:2000:single}, there exists a continuous function $h_c$ such that $h_c(\bx) = c_i$ if $\bx$ in $\cK_i$.
We say that a function $h:\cX\to\bbR$ can separate $m$ disjoint regions $\cK_1, \cK_2,\ldots,\cK_m$ at $\bx_0$ if
$|h_c(\bx_0) - h(\bx_0) | < \delta = \frac{1}{2}\min\{|c_i -c_j| \mid \forall i \neq j\}$.
From result~\ref{theo:scene2:uap}, we can obtain a function $f:\cX\to\bbR$ in the form $\sum_{\bap}w_{\bap}\psi_{\bap}(\bx)$
such that $| h_c(\bx_0) - f(\bx_0) | < \delta = \frac{1}{2}\min\{|c_i -c_j| \mid \forall i \neq j\}$ for all  $\bx_0$ in $\cX$.
Therefore, $f$ can separate $\cK_1, \cK_2,\ldots,\cK_m$. 

We note that the number of observables $O_{\bap}$ in the parallel scenario does not need to scale exponentially with respect to the number of qubits $N$.
From the construction of the circuits, for each $k$ ($1\leq k \leq d$), any combination of $\alpha_{k}, \alpha_{k+d},\alpha_{k+2d},\ldots$ with $p$ nonzero elements gives the same terms in the basis functions $\psi_{\bap}$.
Hence, for each $p$, we only need to choose one combination to construct the observable $O_{\bap}$.
Let $q(k)$ denote the number of values that $p$ can take for each $k$.
Then, the number of observables $O_{\bap}$ does not need to be larger than $q(1)q(2)\ldots q(d)$.
Because the number of elements in $\alpha_{k}, \alpha_{k+d},\alpha_{k+2d},\ldots$ does not exceed $1 + \lfloor\dfrac{N-1}{d} \rfloor$,
the value of $p$ is taken in $0, 1,\ldots, 1 + \lfloor\dfrac{N-1}{d} \rfloor$,
where $\lfloor r \rfloor$ denotes the greatest integer less than or equal to $r$.
Therefore, $q(k)\leq 2 + \lfloor\dfrac{N-1}{d} \rfloor$ for each $k$; thus, the number of observables does not exceed $\left(2 + \lfloor\dfrac{N-1}{d} \rfloor\right)^d$.

Next, we show that the nonlinearity to establish the UAP can be implemented by a special kind of data pre-processing with an activation function incorporated into $\theta_j(\bx)$.
The activation function can be computed by a classical algorithm on the level of logical gates and then translated into a reversible routine to be used as a quantum algorithm~\cite{schuld:2018:bookQML}.
Given an activation function $\sigma: \bbR \to [-1, 1]$, we further assume two conditions for $\sigma$.  
First, $\sigma$ is nonconstant and piecewise continuous.
Here, $\sigma$ is said to be piecewise continuous if it has a finite number of discontinuities in any interval, and its left and right limits are defined (not necessarily equal) at each discontinuity.
Second, $\sigma_{\ba, b}(\bx)=\sigma(\ba\cdot\bx+b)$ is dense in $L^2(\cX)$ where $\ba \cdot \bx$ denotes the inner product of vectors $\ba$ and $\bx$ in $\bbR^d$.
This means that for any $\varepsilon > 0$ and $g \in L^2(\cX)$, there exist $\ba\in \bbR^d$ and $b \in \bbR$ such that $\Vert g - \sigma_{\ba, b}\Vert_{L^2(\cX)} < \varepsilon$.
We apply $\cV_N(\bx)$ in Eq.~\eqref{eqn:feature:v} with 
$
\theta_j(\bx)=\textup{arccos}\left( \sqrt{\dfrac{1+\sigma_{\ba_j, b_j}(\bx)}{2}} \right),
$
where $\ba_j\in\bbR^d$ and $b_j\in \bbR$ are randomly generated from any continuous sampling distribution for each $j$.
In this scheme, the number of observables can be reduced to $N$.
We consider the observables $O_j=I\otimes\ldots\otimes\underbrace{Z}_\text{$j$-index}\otimes\ldots\otimes I$ ($1\leq j\leq N$) with the corresponding basis functions 
\begin{align}\label{eqn:basis:rs2}
     \psi_j(\bx) 
     &= \bra{0}^{\otimes N} \cV_N^{\dagger}(\bx) O_j \cV_N(x) \ket{0}^{\otimes N}\nonumber\\ 
     &= \bra{0} e^{i \theta_j Y} Z e^{-i \theta_j Y} \ket{0} = \sigma_{\ba_j, b_j}(\bx).
\end{align}
Result~\ref{theo:scene1} is obtained from the main result in the UAP of the classical framework in Ref.~\cite{huang:2007:convex} (Theorem 2.3), which states that for any $\varepsilon > 0$ there exist $N$ and $\{w_j\}_{j=1}^N (w_j \in \bbR)$ such that
$
     \Big\Vert \sum_{j=1}^N w_j\sigma_{\ba_j, b_j} - g \Big\Vert_{L^2(\cX)} < \varepsilon.
$

\begin{result}[UAP when implementing activation functions in pre-processing]\label{theo:scene1}
For any continuous function $g: \cX \to \bbR$ and the construction of basis functions $\psi_j$ in Eq.~\eqref{eqn:basis:rs2}; then for any $\varepsilon > 0$, there exist $N$ and $\{w_j\}_{j=1}^N (w_j \in \bbR)$ such that
$
    \Big\Vert \sum_{j=1}^N w_j\psi_j-g\Big\Vert_{L^2(\cX)} < \varepsilon.
$
\end{result}

Result~\ref{theo:scene1} implies that with a sufficient number of qubits, the framework induced from the nonlinear activation function with the selected observables can work as a universal approximator to any continuous function $g:\cX\to\bbR$ in $L^2(\cX)$ with any arbitrary precision.
Similar to the analysis from result~\ref{theo:scene2:uap}, we consider the function $h_c$ to investigate the classification capability in this setting.
From result~\ref{theo:scene1}, for $\varepsilon > 0$, there exists $f:\cX \to \bbR$ in the form of Eq.~\eqref{eqn:linear:qelm} such that $\Big\Vert h_c - f\Big\Vert_{L^2(\cX)} < \varepsilon$.
Let $\cY = \{\by \in \cX \mid | h_c(\by) - f(\by) | \geq \delta\}$ and $V_{\cY}$ be the volume of $\cY$;
we then have $V_{\cY}^{1/2}\delta < \varepsilon$ or $V_{\cY} < (\varepsilon/\delta)^2$.
Therefore, by selecting sufficiently small $\varepsilon$, we can reduce $V_{\cY}$ as small as possible to increase the classification capability.

\textit{Sequential scenario.---} In the parallel scenario, it is assumed that we can increase the number of qubits to approximate the output function to a target continuous function with arbitrary precision.
However, there is a limitation in the current realistic model with a large number of qubits.
We investigate whether the UAP can be obtained by constructing the quantum feature map with only a single qubit by repeating a simple quantum circuit $V(\bx)$ [Fig.~\ref{fig:circuit}(b)].
Unlike the parallel scenario, the quantum feature map described in the following paragraph is not capable of approximating a function whose domain is an infinite set (see \cite{supp}).
We restrict the input set to a finite set $\cX=\{\bx_1, \bx_2, \ldots, \bx_M\}$. 
For example, in a real-world application, $\cX$ can be the set of RGB fixed-size images.

To obtain the UAP, it is important to set the appropriate form of $V(\bx)$.
In \cite{supp}, we present a counter-example of $V(\bx)$ that we cannot obtain the UAP.
Here, we consider the unitary operator $V(\bx)= e^{-\pi i \theta(\bx) Y}$ applied to the single qubit and establish the condition of $\theta(\bx_1), \theta(\bx_2),\ldots,\theta(\bx_M)$ to obtain the UAP.
The quantum feature map is constructed by repeating $V(\bx)$, that is, applying $V^n(\bx) = e^{-n \pi i \theta(\bx) Y}$ ($n \in \mathbb{N}$) to $\ket{0}$, where $\theta:\cX \to \bbR$.
The corresponding basis function with the observable $Z$ (Pauli-Z) becomes
\begin{align}\label{eqn:scene3:basicfunc}
        \psi_n(\bx) &= \bra{0} (V^{n})^{\dagger}(\bx) Z V^n(\bx) \ket{0}= 2 \cos^2 (\pi n \theta(\bx)) - 1 \nonumber\\
    &= \cos(2\pi n\theta(\bx)) = \cos(2 \pi \{n \theta(\bx)\}), 
\end{align}
where $\{n \theta(\bx)\}=n\theta(\bx) - \lfloor n \theta(\bx) \rfloor$ is the fractional part of $n\theta(\bx)$.
The UAP is studied via the Kronecker--Weyl theorem~\cite{king:1994:three,stein:2003:fourier} 
on the density of the fractional parts $(\{n\theta(\bx_1)\}, \ldots,\{n\theta(\bx_M)\})_{n\in\bbN}$. 
In \cite{supp}, we prove the following result, which states that with the condition of the linear independence for $1, \theta(\bx_1), \ldots, \theta(\bx_M)$, any function in $\cX$ can be approximated by repeatedly applying $V(\bx)$ with an appropriate iteration number $n$.
Here, real numbers $b_1, b_2,\ldots,b_L$ are linearly independent over the set of rational numbers $\bbQ$ if the only integral solution to $z_1b_1+z_2b_2+\ldots+z_Lb_L=0$ is the all zero $z_1=z_2=\ldots=z_L=0$.

\begin{result}[UAP in the sequential scenario]\label{theo:3}
If $\cX=\{\bx_1, \bx_2, \ldots, \bx_M\}\subset \bbR^d$ and $1, \theta(\bx_1), \ldots, \theta(\bx_M)$ are linearly independent over $\mathbb{Q}$,
then for any function $g: \cX\to\bbR$ and for any $\varepsilon > 0$, there exist $n \in\mathbb{N}$ and $w\in\bbR$ such that
$
    \left| w\psi_n(\bx) - g(\bx) \right| < \varepsilon 
$
for all $\bx$ in $\cX$. Here, the basis function $\psi_n(\bx)= \cos(2 \pi \{n \theta(\bx)\})$ is defined as in Eq.~\eqref{eqn:scene3:basicfunc}.
\end{result}
Similar to the analysis from result~\ref{theo:scene2:uap}, we can also obtain the classification capability via result~\ref{theo:3}.

\textit{Approximation rate.---} An interesting theoretical question is how to describe relative goodness or badness in a universal approximation.
The approximation rate can be used here, which is the decay rate of the approximation error.
This rate refers to the speed at which the approximation error decreases when the parameters, such as the number of qubits $N$  and the input dimension $d$, are increased.
The approximation rate strongly depends on the nature of the target function $g$ to be approximated and the type of the input set $\cX$.
In \cite{supp}, we prove the following result, which describes the approximation rate in the parallel scenario.
\begin{result}[Approximation rate]\label{theo:4}
If $\cX = [0, 1]^d$ and the target function $g$ is Lipschitz continuous with respect to the Euclidean norm,
we can construct an explicit form of the approximator to $g$ in the parallel scenario by $N$ qubits with the error $\varepsilon=O(d^{7/6}N^{-1/3})$.
Furthermore, we can achieve an approximation error with a better approximation rate in terms of $N$ as
$\varepsilon=O(d^{3/2}N^{-1})$.
\end{result}

The approximation error $\varepsilon=O(d^{3/2}N^{-1})$ can be obtained by using the Jackson theorem of the quantitative information on the degree of polynomial approximation to a continuous function~\cite{newman:1964:jackson}.
It implies that $O(d^{3/2}\varepsilon^{-1})$ qubits are enough to obtain an approximation with $\varepsilon$-error.
However, the explicit form of this approximator remains for future work.

The approximation rate provides a method to compare the asymptotic universality between our quantum feature framework and the classical neural networks.
The number of observables $K$ in our framework corresponds with the number of parameters in the classical neural networks.
Since $K = O(N^d)$ in the parallel scenario, we can write our best approximation error as $\varepsilon = O(K^{-1/d})$ if we fix $d$ and focus on $K$.
Interestingly, this is also the best approximation when using a classical neural network to approximate a  Lipschitz continuous function~\cite{mhaskar:1996:approx,yarotsky:2017:error}.
This result suggests a strong guarantee that the QML models in quantum-enhanced feature spaces can exhibit at least the same expressivity as the classical ML models.

\textit{Conclusion.---}We present a comprehensive understanding of the UAP of quantum feature frameworks induced from quantum-enhanced feature spaces.
This research lays a foundation for further theoretical analysis of the expressivity of these  frameworks and provides insights into the design of a good expressive model in QML applications.
Our proposal addresses the theoretical research question about whether QML models in quantum-enhanced feature spaces can solve the tasks that conventional ML models can in classical settings.
We obtain the results that under typical quantum feature map settings, the QML models can achieve both UAP and classification capability and can thus handle a wide class of ML tasks. 
The suggestions in practical applications are left for future works, such as finding an efficient scheme with the lowest implementation cost to obtain the necessary approximation accuracy.

\begin{acknowledgments}
K.N. and Q.H.T. were supported by MEXT Quantum Leap Flagship Program (MEXT Q-LEAP) Grant Nos. JPMXS0118067394 and JPMXS0120319794.
T. G. and Q. H. T. contributed equally to this work.
\end{acknowledgments} 

\providecommand{\noopsort}[1]{}\providecommand{\singleletter}[1]{#1}%

\end{document}


\title{Supplementary Material for\\
``Universal Approximation Property of Quantum Machine Learning Models in Quantum-Enhanced Feature Spaces''}

\makeatletter
\def\@fnsymbol#1{\ensuremath{\ifcase#1\or \dagger\or * \or \ddagger\or
   \mathsection\or \mathparagraph\or \|\or **\or \dagger\dagger
   \or \ddagger\ddagger \else\@ctrerr\fi}}
   
\def\frontmatter@thefootnote{%
 \altaffilletter@sw{\@fnsymbol}{\@fnsymbol}{\csname c@\@mpfn\endcsname}%
}%
\makeatother

\author{Takahiro Goto}
\email{goto.takahiro.2020@gmail.com}
\affiliation{
    Reservoir Computing Seminar Group, 
    Nagase Hongo Building F8, 5-24-5, Hongo, Bunkyo-ku, Tokyo 113-0033, Japan
}

\author{Quoc Hoan Tran}
\email{tran\_qh@ai.u-tokyo.ac.jp (Corresponding author, equally contributed with T. Goto)}

\author{Kohei Nakajima}
\email{k\_nakajima@mech.t.u-tokyo.ac.jp}
\affiliation{
    Reservoir Computing Seminar Group, 
    Nagase Hongo Building F8, 5-24-5, Hongo, Bunkyo-ku, Tokyo 113-0033, Japan
}
\affiliation{
	Graduate School of Information Science and Technology, The University of Tokyo, Tokyo 113-8656, Japan
}

\date{\today}

\begin{abstract}
This supplementary material describes in detail the calculations introduced in the main text.
The equation numbers in this section are
prefixed with S (e.g., Eq.\textbf{~}(S1)),
while numbers without the prefix (e.g., Eq.~(1)) refer to items in the main text.
\end{abstract}

\maketitle
\tableofcontents
\newpage
\section{Proof of Result 1}

\boxedtitledtext{RESULT 1 (UAP in the parallel scenario)}{
For any continuous function $g: \cX \to \bbR$; 
then for any $\varepsilon>0$, there exist $N$ and a collection of output weights $w_{\bap}$ and observables $O_{\bap} = Z^{\alpha_1}\otimes Z^{\alpha_2}\otimes \ldots \otimes Z^{\alpha_N}$, where $\bap = (\alpha_1, \alpha_2, \ldots, \alpha_N) \in \{0, 1\}^N$ such that
  $
    | \sum_{\bap} w_{\bap} \psi_{\bap}(\bx)  - g(\bx) | < \varepsilon
  $
  for all $\bx$ in $\cX$.
}

\paragraph{Proof.---}
First, we prove the following lemma~\ref{lemma:S1} to show that any polynomial function in $\cX$ can be constructed from basis functions $\psi_{\bap}(\bx)$ mentioned in the parallel scenario.

\begin{lemma}\label{lemma:S1}
Consider the polynomial $P(\bx)$ of the input $\bx= (x_1, x_2, \ldots, x_d) \in [0, 1]^d$
where the degree of $x_j$ in $P(\bx)$ is less than or equal to $\dfrac{N + (d - j)}{d}$ for $j=1,\ldots,d$ $(N\geq d)$; there then exists a collection of output weights $\{w_{\bap} \in \bbR \mid \bap \in \{0, 1\}^N\}$
  such that 
  \begin{align}\nonumber
      \sum_{\bap \in \{0, 1\}^N} w_{\bap} \psi_{\bap}(\bx) = P(\bx).
  \end{align}
\end{lemma}

Consider a real number $r$; we denote $\lfloor r \rfloor$ as the greatest integer less than or equal to $r$ and $\lceil r \rceil$ as the least integer greater than or equal to $r$.
For integers $i, d$ $(d>0)$, let $[i]$ denote the integer number $k$ such that $1\leq k \leq d$ and $i \equiv k (\textup{mod } d)$.
For a nonzero monomial $m(\bx) = l x_1^{a_1} x_2^{a_2} \cdots x_d^{a_d}$, let $\deg(m(\bx)) = a_1 + a_2 + \cdots a_d$ be the degree of $m(\bx)$. We define $\deg(0)=-1$ for our convenience.
Furthermore, for a nonzero polynomonial $P(\bx)$, the degree $\deg(P(\bx))$ of $P(\bx)$
is defined as the largest degree of monomial terms in $P(\bx)$.
  
The basis functions induced from  $
    \cV_N(\bx) = V_1(\bx) \otimes V_2(\bx) \otimes \ldots \otimes V_N(\bx)
    $
and $
    O_{\bap} = Z^{\alpha_1}\otimes Z^{\alpha_2}\otimes \ldots \otimes Z^{\alpha_N},
$
where
$
    V_j(\bx) = e^{-i \textup{arccos}(\sqrt{x_{[j]}})Y}, (1\leq j \leq N)
$ and $\bap = (\alpha_1, \alpha_2, \ldots, \alpha_N) \in \{0, 1\}^N$, are represented as follows:
  \begin{align}
    \psi_{\bap} (\bx)  \nonumber 
    &= \bra{0}^{\otimes N} \cV_N^{\dagger}(\bx) O_{\bap} \cV_N(\bx) \ket{0}^{\otimes N} \\ \nonumber
    &= \braket{0 |^{\otimes N} \left(V_1(\bx) \otimes V_2(\bx) \otimes \ldots \otimes V_N(\bx)\right)^{\dagger}  
                                  \left(Z^{\alpha_1} \otimes Z^{\alpha_2} \otimes \cdots \otimes Z^{\alpha_N} \right)
                                  \left(V_1(\bx) \otimes V_2(\bx) \otimes \ldots \otimes V_N(\bx)\right) |     0}^{\otimes N} \\ \nonumber
    &= \prod_{i=1}^{N} \braket{0 | V_{i}^{\dagger}(\bx) Z_i^{\alpha_i} V_{i}(\bx) | 0} \\ \nonumber
    &= \prod_{i=1}^{N} \left(2 \left(\arccos \left( \cos \sqrt{x_{[i]}}\right) \right)^2 -1 \right)^{\alpha_i} \\ 
    &= \prod_{i=1}^{N} (2 x_{[i]} - 1)^{\alpha_i} = 2^{|\bap|} x_{[1]}^{\alpha_1} x_{[2]}^{\alpha_2} \cdots x_{[N]}^{\alpha_N} + Q_{\bap}(\bx).\label{eqn:basic:form}
  \end{align}
Here, $|\bap| = \alpha_1 + \alpha_2 + \ldots + \alpha_N$, and $Q_{\bap}(\bx)$ is the total of other monomial terms in $\psi_{\bap}(\bx)$ with the degree less than $|\bap|$.

For each $j$ $(1 \leq j \leq d)$, we find the number of integer $i$ such that $1\leq i \leq N$ and $[i] \equiv j (\textup{mod } d)$. It is equivalent to finding the number of integer $p$ such that $1 -j \leq dp \leq N - j$ (then $i = j + dp$). Because $1-j \geq 1 - d > -d$, then $0\leq dp < N- j + 1$ or $0\leq p < \dfrac{N-j+1}{d}$. Therefore, the number of such $p$ is $\lceil \dfrac{N-j+1}{d} \rceil$. From this result and Eq.~\eqref{eqn:basic:form}, the degree of $x_j$ ($1\leq j \leq d$) in $\psi_{\bap}(\bx)$ is not larger than $\lceil \dfrac{N-j+1}{d} \rceil = \lfloor \dfrac{N + (d - j)}{d} \rfloor$.

Next, to prove lemma S1, we prove that if $P(\bx)$ is the polynomial where the degree of $x_j$ in $P(\bx)$ is less than or equal to $\lfloor \dfrac{N + (d - j)}{d} \rfloor$, then $P(\bx)$ can be represented as the form $\sum_{\bap}w_{\bap}\psi_{\bap}(\bx)$. 
The proof is by induction on the degree of $P(\bx)$.
The statement is trivial for the cases $\deg(P(\bx))=-1$ and $\deg(P(\bx))=0$. We assume that the statement is established if $\deg(P(\bx))=-1, 0, \ldots, D-1$ (induction assumption); then we need to establish the truth of the statement for the case $\deg(P(\bx)) = D > 0$ (induction hypothesis).
Let $m_1(\bx), m_2(\bx), \ldots, m_K(\bx)$ denote all monomial terms in $P(\bx)$ such that $\deg(m_k)(\bx)=\deg(P(\bx))$ for all $k$ ($1\leq k \leq K$). Then, $P(\bx)$ can be represented as
\begin{align}
    P(\bx) = m_1(\bx) + m_2(\bx) + \ldots + m_K(\bx) + P^{\prime}(\bx),
\end{align}
where $\deg(P^{\prime}(\bx)) < \deg(P(\bx))$.

For each $k$ ($1\leq k \leq K$), let $\bap^{(k)}$ denote the value of $\bap$ such that $m_k(\bx)=w_{\bap}2^{|\bap|} x_{[1]}^{\alpha_1} x_{[2]}^{\alpha_2} \cdots x_{[N]}^{\alpha_N}$.
We consider the difference polynomial $R(\bx)$ between $P(\bx)$ and $\sum_{k=1}^Kw_{\bap^{(k)}}\psi_{\bap^{(k)}}(\bx)$ as
  \begin{align}\label{eqn:rx:remainder}
    R(\bx) = P(\bx) - \sum_{k=1}^{K} w_{\bap^{(k)}} \psi_{\bap^{(k)}} (\bx) = P'(\bx) - \sum_{k=1}^{K} w_{\bap^{(k)}} Q_{\bap^{(k)}}(\bx).
  \end{align}
 Because $\deg(R(\bx)) < \deg(P(\bx))=D$ and the degree of $x_j$ in $R(\bx)$ is not larger than $\lfloor \dfrac{N + (d - j)}{d} \rfloor$, following the induction assumption, $R(\bx)$ can be written in the form $\sum_{\bap}w_{\bap}\psi_{\bap}(\bx)$.
 From this result and Eq.~\eqref{eqn:rx:remainder}, $P(\bx)$ is also represented by the form $\sum_{\bap}w_{\bap}\psi_{\bap}(\bx)$. Therefore, the induction hypothesis is proved for $\deg(P(\bx))=D$. The statement in lemma~\ref{lemma:S1} is established for all values of $\deg(P(\bx))$.

Result 1 is obtained from lemma~\ref{lemma:S1} and the Weierstrass' Polynomial Approximation Theorem, which states that any continuous function on $\cX$ can be approximated by polynomial functions with arbitrary precision in terms of the supremum norm. Let us mention here the general form of the Weierstrass' Polynomial Approximation Theorem, which is known as the Stone--Weierstrass theorem. The statement of theorem~\ref{theo:stone} is directly quoted from Ref.~\cite{yoshida:1980:functional}.

  \begin{theorem}[Stone--Weierstrass~\cite{yoshida:1980:functional}]\label{theo:stone}
    Let $\cX$ be a compact space and $C(\cX)$ be the set of real-valued continuous functions defined on $\cX$. Let a subset $B$ of $C(\cX)$ satisfy the three conditions: 
    \begin{enumerate}[(i)]
        \item If $f, g \in B$, then the product $f \cdot g$ and linear combination $\alpha f + \beta g$, with real coefficients $\alpha, \beta$, belong to B.
        \item The constant function $1$ belongs to $B$.
        \item The uniform limit $f_\infty$ of any sequence ${f_n}$ of functions in $B$ also belongs to $B$.
    \end{enumerate}
     Then $B=C(\cX)$ iff $B$ separates the points of $\cX$, i.e. iff, for every distinct points $\bx, \by$ in $\cX$, there exists a function $h$ in $B$ such that $h(\bx) \neq h(\by)$.
  \end{theorem}

Let $\cP$ denote the set of all polynomials on the compact set $\cX \subset \bbR^d$.
We apply theorem~\ref{theo:stone} for the uniform closure $\bar{\cP}$, which is the space of all functions that can be approximated by a sequence of uniformly converging polynomials in $\cP$.
We check three conditions given in theorem~\ref{theo:stone}. From the definition of $\bar{\cP}$, the conditions $(ii)$ and $(iii)$ are satisfied. Let us check the condition $(i)$. For arbitrary functions $f, g \in \bar{\cP}$, there exist sequences of polynomials $\{f_n\}, \{g_n\} \in \cP$, such that $\lim_{n\to \infty} \Vert f_n - f \Vert_{\infty} = 0$ and  $\lim_{n\to \infty} \Vert g_n - g \Vert_{\infty} = 0$. We define two functions, $p=f\cdot g$ and $q=\alpha f + \beta g$, and two sequences, $\{p_n\}$ and $\{q_n\}$, where $p_n=f_n\cdot g_n$ and $q_n=\alpha f_n + \beta g_n$.
Because $f$ and $g$ are defined on the compact set $\cX$, there are the maximum values and minimum values for $f$ and $g$. Let $M$ be a real number such that $\Vert f \Vert_{\infty} \leq M$ and $\Vert g \Vert_{\infty} \leq M$, and then
\begin{align}\label{eqn:fg}
    \Vert p_n - p\Vert_{\infty} &= \Vert f_n\cdot g_n - f\cdot g\Vert_{\infty} \nonumber\\
    &= \Vert (f_n - f)\cdot g +  (g_n - g) \cdot f_n \Vert_{\infty}\nonumber\\
    &\leq \Vert f_n - f \Vert_{\infty}  \Vert g \Vert_{\infty} + \Vert g_n - g \Vert_{\infty} \Vert f_n \Vert_{\infty}\nonumber\\
    &\leq M \Vert f_n - f \Vert_{\infty} + M \Vert g_n - g \Vert_{\infty}.
\end{align}
Similarly, we have the following inequality
\begin{align}\label{eqn:f+g}
    \Vert q_n - q\Vert_{\infty} &= \Vert (\alpha f_n + \beta g_n) - (\alpha f + \beta g) \Vert_{\infty}\nonumber\\
    &= \Vert \alpha (f_n - f) + \beta (g_n-g)\Vert_{\infty}\nonumber\\
    &\leq |\alpha| \Vert f_n - f \Vert_{\infty} + |\beta| \Vert g_n - g \Vert_{\infty}.
\end{align}
Equations~\eqref{eqn:fg} and \eqref{eqn:f+g} imply that $\lim_{n\to \infty} \Vert p_n - p \Vert_{\infty} = 0$ and  $\lim_{n\to \infty} \Vert q_n - q \Vert_{\infty} = 0$, and therefore $p\in\cP$ and $q\in\cP$; the condition $(i)$ is satisfied.

Furthermore, we consider arbitrary distinct points $\bx=(x_1,\ldots,x_d)$ and $\by=(y_1,\ldots,y_d) \in \cX$. 
Without a loss of generality, we assume that $x_1 \neq y_1$. We consider a polynomial $h$ in $\cP$ ($h \in \bar{\cP}$) such that $h(\bz)=z_1 - x_1$ for all $\bz=(z_1,\ldots,z_d) \in \cX$.
Then, $h(\bx) = 0$ and $h(\by)=y_1-x_1 \neq 0$.
Therefore, $\bar{\cP}$  can separate the points of $\cX$. From the Stone--Weierstrass theorem, we obtain $\bar{\cP} = C(\cX)$, or $\cP$ is dense in $C(\cX)$. 

Now, let $g$ be a continuous real-valued function defined on $\cX$. Because $\cP$ is dense in $C(\cX)$, there exists a sequence of polynomials $P_k(\bx)\in\cP$ such that $\lim_{k\to\infty}\Vert g(\bx)-P_k(\bx)\Vert_{\infty}=0$. Then, for any $\varepsilon>0$, there exists $k_0$ such that $\Vert P_k(\bx) - g(\bx)\Vert_{\infty} < \varepsilon$ for all $k\geq k_0$. We choose the number $N_{k_0}$ of qubits such that the degree of $x_j$ in $P_{k_0}(\bx)$ is less than or equal to $\dfrac{N+(d-j)}{d}$ for all $N\geq N_{k_0}$. From lemma~\ref{lemma:S1}, for $N\geq N_{k_0}$, there exists a collection of output weights $\{w_{\bap} \in \bbR \mid \bap \in \{0, 1\}^{N}\}$
  such that 
  $
      \sum_{\bap \in \{0, 1\}^{N}} w_{\bap} \psi_{\bap}(\bx) = P_k(\bx)
  $. Then,
  \begin{align}
      \Vert \sum_{\bap \in \{0, 1\}^{N}} w_{\bap} \psi_{\bap}(\bx) - g(\bx)\Vert_{\infty} < \varepsilon
  \end{align}
  for $N\geq N_{k_0}$. Therefore, we have the following result, which is Result 1 in the main text:
  \begin{align}\nonumber
    \lim_{N \to \infty} \min_{\bw=\{\bw_{\bap}\} \in \mathbb{R}^{2^n}} \Big\Vert  \sum_{\bap \in \{0, 1\}^N} w_{\bap} \psi_{\bap}(\bx) - g(\bx)\Big\Vert_{\infty} = 0.
  \end{align}

\section{A counter-example of the sequential scenario without UAP}
\begin{figure}
 	\includegraphics[width=9cm]{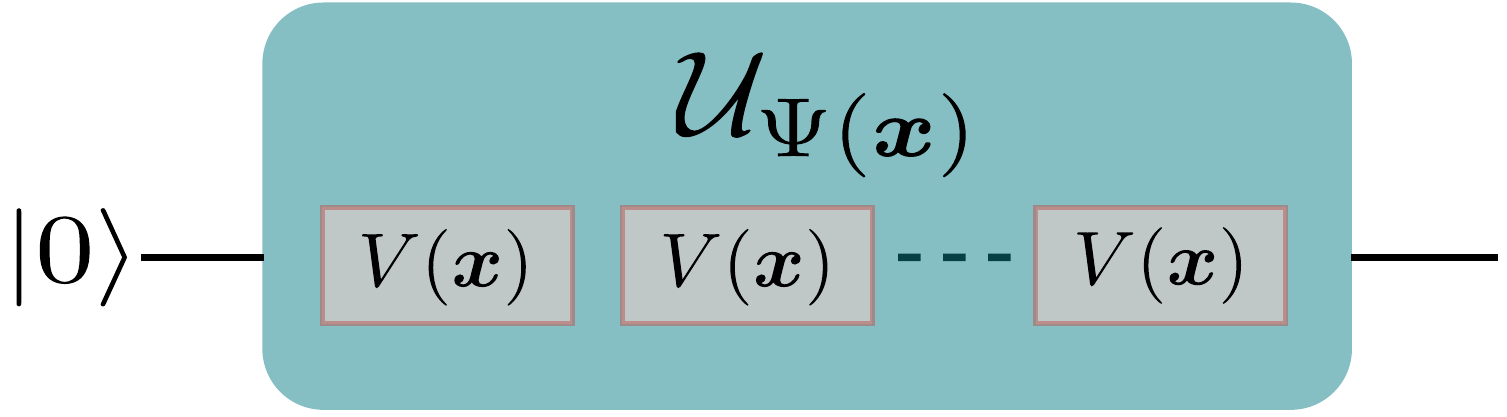}
	\protect\caption{The sequential scenario in setting the quantum feature map. The circuit is the repetition of a simple circuit $V(\bx)$ (for example, a single Pauli-Y rotation) acting on the same qubits.
 		\label{fig:seq}}
\end{figure}

In the sequential scenario (Fig.~\ref{fig:seq}), we investigate whether the UAP can be obtained by constructing the quantum feature map with only a single qubit by repeating the same randomness-free and simple quantum circuit $V(\bx)$.
To obtain the UAP, it is important to set the appropriate form of $V(\bx)$.

We give here a counter-example of $V(\bx)$ that we cannot obtain the UAP. For example, if $V(\bx)$ is a unitary operator such that $\ket{0}$ is a common eigenvector of $V(\bx_1)$ and $V(\bx_2)$,
we denote $\lambda_1$ and $\lambda_2$ as the corresponding eigenvalues of $V(\bx_1)$ and $V(\bx_2)$, respectively. Then the quantum states at $\bx_1, \bx_2$ after repeating $V(\bx)$ for $n$ times are
  \begin{align}
    \ket{\Psi(\bx_1)} &= V^n(\bx_1) \ket{0} = \lambda_{1}^n \ket{0},\\
    \ket{\Psi(\bx_2)} &= V^n(\bx_2) \ket{0} = \lambda_{2}^n \ket{0}.
  \end{align}
  For an arbitrary set of observable operator $\bO = \{O_1, O_2, \ldots, O_K\}$, the values at $\bx_1, \bx_2$ of the basis function $\psi_{i}(\bx)$ corresponding with the observable $O_i$ are
  \begin{align}
      \psi_{i}(\bx_1) &= \bra{\Psi(\bx_1)} O_i \ket {\Psi(\bx_1)} = |\lambda_{1}|^{2n} \bra{0} O_i \ket{0},\\
      \psi_{i}(\bx_2) &= \bra{\Psi(\bx_2)} O_i \ket {\Psi(\bx_2)} = |\lambda_{2}|^{2n} \bra{0} O_i \ket{0}.
  \end{align}
  Since $V(\bx_1)$ and $V(\bx_2)$ are unitary, 
  then $|\lambda_1|=|\lambda_2| = 1$ and $\psi_i(\bx_1) = \psi_i(\bx_2)$ for all $i$. 
  Hence, any linear combination of $\{\psi_i(\bx)\}$ cannot approximate the function which takes different values at $\bx_1$ and $\bx_2$.
  Therefore, we cannot obtain the UAP for this setting of $V(\bx)$.

\section{UAP of the sequential scenario on an infinite set}
In the main text, we consider the following form of $V(\bx)$ as $V(\bx)= e^{-\pi i \theta(\bx) Y}$, where $\theta:\cX \to \bbR$. This unitary operator is applied to the single qubit.
The quantum feature map is constructed by repeating $V(\bx)$, that is, applying $V^n(\bx) = e^{-n \pi i \theta(\bx) Y}$ ($n \in \mathbb{N}$) to $\ket{0}$.
We consider the observable $Z$ (Pauli-Z) on the state after applying the quantum feature map. Therefore, the corresponding basis function is
\begin{align}\label{eqn:scene3:basicfunc}
        \psi_n(\bx) &= \bra{0} (V^{n})^{\dagger}(\bx) Z V^n(\bx) \ket{0}\\
    &= 2 \cos^2 (\pi n \theta(\bx)) - 1 = \cos (2 \pi \{n \theta(\bx)\})\nonumber, 
\end{align}
where $\{n \theta(\bx)\}=n\theta(\bx) - \lfloor n \theta(\bx) \rfloor$ is the fractional part of $n\theta(\bx)$.
The output function is modeled by the output weight $w \in \bbR$ as
$
    f_n(\bx; w) = w\psi_n(\bx) = w \left(\cos(2 \pi \{n \theta(\bx)\}) - 1\right).
$

We prove that the quantum feature map described above is not capable of approximating a function whose domain is an infinite set.
Here, for simplicity, we consider $\cX$ as a closed interval $\cX =[a, b] \subset \bbR$, where $b > a$, and $g$ is an arbitrary continuous real-valued function defined on $[a, b]$. We assume that $\theta$ is a continuous real-valued function defined on $[a, b]$.
If $\theta(a) = \theta(b)$, any basis function $\psi$ constructed from any observable takes the same value on $a$ and $b$, that is, $\psi(a)=\psi(b)$; therefore, the linear combination of these basis functions cannot approximate $g$ if $g(a) \neq g(b)$. Therefore, we only need to consider the case $\theta(a)\neq\theta(b)$ to investigate the UAP.
Without a loss of generality, we assume that $\theta(a) < \theta(b)$. Because the set of rational numbers is a dense subset of the real numbers, there exist rational numbers $q_1, q_2 \in \bbQ$ such that $\theta(a) < q_1 < q_2 < \theta(b)$. From the intermediate value theorem, there exist $c_1, c_2 \in [a, b]$ such that $\theta(c_1)=q_1, \theta(c_2)=q_2$. If we write $q_1=n_1/s, q_2=n_2/s$ for $n_1, n_2, s\in \mathbb{Z}$, then the fractional sequence $(\{n\theta(c_1)\}, \{n\theta(c_2)\})_{n\in\mathbb{N}}$ has the period $s$ and takes the finite number of values.
Therefore, $(\psi_n(c_1), \psi_n(c_2))_{n\in\mathbb{N}}$ also takes the finite number of values.

We assume that the quantum feature framework defined by this setting has the UAP. Let us choose a continuous function $g: [a, b]\to\bbR$ such that $g(c_1)=kg(c_2) > 0$ for $k>1$. For $\varepsilon_k = g(c_2)/k$, there exist $w_k\in\bbR$ and $n_k\in\bbN$ such that $|f_{n_k}(c_1; w_k) - g(c_1)| < \varepsilon_k$ and $|f_{n_k}(c_2; w_k) - g(c_2)| < \varepsilon_k$.
Therefore, we have the following inequalities:
\begin{align}
    g(c_1)-\varepsilon_k &< f_{n_k}(c_1; w_k) = w_k\psi_{n_k}(c_1) < g(c_1) + \varepsilon_k, \\
    g(c_2)-\varepsilon_k &< f_{n_k}(c_2; w_k) = w_k\psi_{n_k}(c_2) < g(c_2) + \varepsilon_k. 
\end{align}
Because $g(c_1)-\varepsilon_k=(k-\dfrac{1}{k})g(c_2) > 0$, $g(c_2)-\varepsilon_k = (1-\dfrac{1}{k})g(c_2) > 0$, 
and $g(c_2)+\varepsilon_k = (1+\dfrac{1}{k})g(c_2) > 0$, 
we have
\begin{align}
    \dfrac{\psi_{n_k}(c_1)}{\psi_{n_k}(c_2)} = \dfrac{w_k\psi_{n_k}(c_1)}{w_k\psi_{n_k}(c_2)}> \dfrac{g(c_1)-\varepsilon_k}{g(c_2) + \varepsilon_k} = \dfrac{k^2-1}{k+1} = k - 1.
\end{align}
Therefore, $\dfrac{\psi_{n_k}(c_1)}{\psi_{n_k}(c_2)}$ can take on an infinite number of values as $k$ varies ($k>1$).
However, this is impossible because $(\psi_n(c_1), \psi_n(c_2))_{n\in\mathbb{N}}$ can only take on a finite number of values.
In other words, we were wrong to assume the quantum feature framework defined by this setting has the UAP. 
We can conclude that the setting of the sequential scenario mentioned in the main script cannot provide the UAP on an infinite set $[a, b]$.
The same statement can also be obtained if the input set $\cX$ is a connected infinite subset of $\bbR^d$.

\section{Proof of Result 3}

Similar to the previous section, we consider the unitary operator $V(\bx)= e^{-\pi i \theta(\bx) Y}$ applied to the single qubit and establish the condition of $\theta(\bx_1), \theta(\bx_2),\ldots,\theta(\bx_M)$ to obtain the UAP.
The quantum feature map is constructed by repeating $V(\bx)$, that is, applying $V^n(\bx) = e^{-n \pi i \theta(\bx) Y}$ ($n \in \mathbb{N}$) to $\ket{0}$, where $\theta:\cX \to \bbR$.
The output function is modeled by the output weight $w \in \bbR$ as
$
    f_n(\bx; w) = w\psi_n(\bx) = w \cos(2 \pi \{n \theta(\bx)\}).
$
The UAP is studied via the density of the fractional parts $(\{n\theta(\bx_1)\}, \ldots,\{n\theta(\bx_M)\})_{n\in\bbN}$. 
We present here the following lemma, which can be directly derived from the Kronecker--Weyl theorem~\cite{king:1994:three,stein:2003:fourier}.

\begin{lemma}\label{lemma:wyel}
If real numbers $1, a_1, a_2,\ldots, a_M$ are linearly independent over the field $\bbQ$ of rational numbers, then $(\{na_1\}, \ldots,\{na_M\})_{n\in\bbN}$ is dense in $[0,1)^M$. That is, given $(t_1,t_2,\ldots,t_M) \in [0,1)^M$, for every $\varepsilon > 0$, there exist $n\in\bbN$ such that $|\{na_i\} - t_i| < \varepsilon$ for  $i=1,2,\ldots,M$.
\end{lemma}

Here, real numbers $b_1, b_2,\ldots,b_L$ are linearly independent over $\bbQ$ if the only integral solution to $\alpha_1b_1+\alpha_2b_2+\ldots+\alpha_Lb_M=0$ is the all zero $\alpha_1=\alpha_2=\ldots=\alpha_L=0$.
Now, we consider the case when $1, \theta(\bx_1), \ldots,\theta(\bx_M)$ are linearly independent over $\bbQ$.
Given a real-valued function $g$ on $\cX$, we define real numbers $\beta_1, \ldots,\beta_M$, such that $\beta_i=g(\bx_i)/\beta$, where $\beta=1 + \max_{i=1,\ldots,M}|g(\bx_i)|$. 
Because $-1<\beta_i<1$, there exists $\gamma_i\in(0, 1)$ such that $\beta_i=\cos(2\pi\gamma_i)$ for $i=1,\ldots,M$.
From lemma~\ref{lemma:wyel}, for every $\varepsilon > 0$, there exists $n\in\bbN$ such that $|\{n\theta(\bx_i)\}-\gamma_i| < \varepsilon/(2\pi\beta)$ for $i=1,\ldots,M$.
If we choose $w=\beta$, then $|f_n(\bx_i;w) - g(\bx_i)|=\beta|\psi_n(\bx_i) - \beta_i|=\beta|\cos(2\pi\{n\theta(\bx_i)\}) - \cos(2\pi\gamma_i)|\leq 2\pi\beta|\{n\theta(\bx_i)\}-\gamma_i| < \varepsilon$ for $i=1,\ldots,M$. We obtain Result 3, which states that with the condition of the linear
independence for $1, \theta(\bx_2), \ldots, \theta(\bx_M)$, any function in $\cX$ can be approximated by repeatedly applying $V(\bx)$ with an appropriate iteration number $n$.

\boxedtitledtext{RESULT 3 (UAP in the sequential scenario)}{
If $\cX=\{\bx_1, \bx_2, \ldots, \bx_M\}\subset \bbR^d$ and $1, \theta(\bx_1), \ldots, \theta(\bx_M)$ are linearly independent over $\mathbb{Q}$, 
then for any function $g: \cX\to\bbR$ and for any $\epsilon > 0$, there exist $n \in\mathbb{N}$ and $w\in\bbR$ such that
$
    \left| f_n(\bx; w) - g(\bx) \right| < \epsilon 
$
for all $\bx$ in $\cX$.
}



Finally, to complete our proof, we prove lemma~\ref{lemma:wyel}.
We present the proof from scratch. 
This lemma can also be directly derived from the Kronecker--Weyl theorem~\cite{king:1994:three}.
The proof for the special case when $M=1$ (which is known as Weyl's equidistribution theorem) can be found in some analysis textbooks, for example, in Ref.~\cite{stein:2003:fourier}.

Let $\ba=(a_1,a_2,\ldots,a_M)\in\bbR^M$, $\bk=(k_1, k_2, \ldots,k_M)\in \bbZ^M$, $\boldsymbol{0}=(0,0,\ldots,0)\in\bbZ^M$ and $\bk \cdot \ba$ denote the dot product between $\bk$ and $\ba$ as $\bk\cdot\ba=k_1a_1+k_2a_2+\ldots+k_Ma_M$.
From the linearly independent property over $\bbQ$ of $1, a_1, a_2, \ldots, a_M$, we have
  \begin{align}
    \et^{2\pi i \bk \cdot \ba} = 1 \Leftrightarrow \bk \cdot \ba \in \bbZ \Leftrightarrow k_1 = k_2=\ldots=k_M=0.
  \end{align}
  Therefore,
  \begin{align}
  \dfrac{1}{M} \sum_{n=1}^{M} \et^{2\pi i n \bk \cdot \ba} = \begin{cases}
    \dfrac{ \et^{2\pi i \bk \cdot \ba} } {M} \dfrac{1 - \et^{2\pi i M \bk \cdot \ba}}{1 - \et^{2\pi i \bk \cdot \ba}} \to 0 \quad (\mathrm{ as }\ M \to \infty), & \text{if $\bk\neq\boldsymbol{0}$}.\\
    1, & \text{if $\bk=\boldsymbol{0}$}.
  \end{cases}
  \end{align}
  Then,
  \begin{equation}\label{eqn:exp}
    \lim_{M \to \infty} \frac{1}{M} \sum_{n=1}^{M} \et^{2\pi i n \bk \cdot \ba} = \int_{\mathbb{T}^M} \et^{2\pi i \bk \cdot \bx} d\bx,
  \end{equation}
  where $\bbT=[0, 1), \bbT^M=[0, 1)^M$.
  Let us consider a trigonometric polynomial $p: \bbT^M \to \mathbb{C}$,
  \begin{align}
      p(\bx) = \sum_{\bk \in \cI}\hat{p}_{\bk}\et^{2\pi i\bk\cdot\bx}, \hat{p}_{\bk}\in\mathbb{C}, \cI\subset\bbZ^M, |\cI| < \infty,
  \end{align}
  supported on an arbitrary frequency index set $\cI$ of finite cardinality. From~\eqref{eqn:exp}, we obtain
  \begin{equation}\label{eqn:poly}
    \lim_{M \to \infty} \frac{1}{M} \sum_{n=1}^{M} p(n\ba) = \int_{\mathbb{T}^M} p(\bx) d\bx.
  \end{equation}

  We apply the Weierstrass' Polynomial Approximation Theorem for the case of trigonometric polynomials~\cite{yoshida:1980:functional}. For an arbitrary continuous real-valued function $f$ defined on $\bbT^M$, there exists a sequence of trigonometric polynomial $\{p_j(\bx)\}_{j\in\bbN}$ uniformly converging to $f(\bx)$. This means that for every $\varepsilon > 0$, there exists a $j_0\in\bbN$ such that $\Vert f - p_j \Vert_{\infty} < \varepsilon$ for all $j\geq j_0$. Therefore, for $j\geq j_0$, we have
  
  \begin{align}
     & \left| \frac{1}{M} \sum_{n=1}^{M} f(n \ba) - \int_{\mathbb{T}^M} f(\bx) d\bx \right| \\
    = &  \left| \frac{1}{M} \sum_{n=1}^{M} f(n \ba) - \frac{1}{M} \sum_{n=1}^{M} p_j(n \ba) \right.
    + \frac{1}{M} \sum_{n=1}^{M} p_j(n \ba) - \int_{\mathbb{T}^M} p_j(\bx) d\bx
    + \left. \int_{\mathbb{T}^M} p_j(\bx) d\bx - \int_{\mathbb{T}^M} f(\bx) d\bx \right|\\
    \leq &  \left| \frac{1}{M} \sum_{n=1}^{M} f(n \ba) - \frac{1}{M} \sum_{n=1}^{M} p_j(n \ba) \right|
    + \left| \frac{1}{M} \sum_{n=1}^{M} p_j(n \ba) - \int_{\mathbb{T}^M} p_j(\bx) d\bx \right|
    + \left| \int_{\mathbb{T}^M} p_j(\bx) d\bx - \int_{\mathbb{T}^M} f(\bx) d\bx \right|\\
    \leq & \frac{1}{M} \sum_{n=1}^{M} ||f - p_j||
    +  \left| \frac{1}{M} \sum_{n=1}^{M} p_j(n \ba) - \int_{\mathbb{T}^M} p_j(\bx) d\bx \right|
    +  \int_{\mathbb{T}^M} ||p_j- f|| d\bx \\
    \leq & 2 \epsilon + \left| \frac{1}{M} \sum_{n=1}^{M} p_j(n \ba) - \int_{\mathbb{T}^M} p_j(\bx) d\bx \right|.\label{eqn:fp}
  \end{align}
From Eq.~\eqref{eqn:poly} and Eq.~\eqref{eqn:fp}, we have $\lim_{M\to\infty}\left| \frac{1}{M} \sum_{n=1}^{M} f(n \ba) - \int_{\mathbb{T}^M} f(\bx) d\bx \right|\leq 2\varepsilon$ for all $\varepsilon > 0$. Then, we have
  \begin{equation}\label{eqn:cont}
    \lim_{M \to \infty} \frac{1}{M} \sum_{n=1}^{M} f(n\ba) = \int_{\mathbb{T}^M} f(\bx) d\bx.
  \end{equation}
Let us consider a distance $\rho$ on $\bbT^M$, and $B(\by, r)=\{\bx\in\bbT^M \mid \rho(\by, \bx) \leq r\}$ for each $\by\in\bbT^M$ and $r>0$. We define the characteristic function of $B(\by, r)$ as
\begin{equation}
    \chi_{B(\by, r)}(\bx) = \begin{cases} 1 & \textup{ if }\rho(\by, \bx) \leq r 
    \\ 0 & \textup{ otherwise. }\end{cases}
\end{equation}
  For any $\epsilon > 0$, we define the following continuous real-valued functions on $\bbT^M$: 
  \begin{equation}
    f_{+}^{(\epsilon)}(\bx) = \begin{cases} 1 & \textup{ if } \rho(\by, \bx) \leq r 
    \\ 1 - \dfrac{\rho(\by, \bx) - r}{\epsilon} & \textup{ if } r < \rho(\by, \bx) \leq r + \epsilon 
    \\  0 & \textup{ otherwise, } \end{cases} 
  \end{equation}
  \begin{equation}
    f_{-}^{(\epsilon)}(\bx) = \begin{cases} 1 & \textup{ if } \rho(\by, \bx) \leq r - \epsilon \\ \dfrac{r - \rho(\by, \bx)}{\epsilon} & \textup{ if } r - \epsilon < \rho(\by, \bx) \leq r \\  0 & \textup{ otherwise.} \end{cases}
  \end{equation}
  Then,
  \begin{equation}\label{eqn:inq}
    \frac{1}{M} \sum_{n=1}^{M} f_{-}^{(\epsilon)}(n \ba) \leq  \frac{1}{M} \sum_{n=1}^{M}\chi_{B(\by, r)}(n \ba) \leq \frac{1}{M} \sum_{n=1}^{M} f_{+}^{(\epsilon)}(n \ba).
  \end{equation}
  Because $f_{+}^{(\epsilon)}$ and $f_{-}^{(\epsilon)}$ are continuous, from Eq.~\eqref{eqn:cont}, we have
  \begin{align}
    \lim_{M \to \infty} \frac{1}{M} \sum_{n=1}^{M} f_{-}^{(\epsilon)}(n \ba) &= \int_{\mathbb{T}^M} f_{-}^{(\epsilon)}(\bx) d\bx\label{eqn:lim1},\\
    \lim_{M \to \infty} \frac{1}{M} \sum_{n=1}^{M} f_{+}^{(\epsilon)}(n \ba) &= \int_{\mathbb{T}^M} f_{+}^{(\epsilon)}(\bx) d\bx\label{eqn:lim2}.
  \end{align}
  Therefore, from Eq.~\eqref{eqn:inq} and Eqs.~\eqref{eqn:lim1} and \eqref{eqn:lim2}, we obtain the following relationships:
  \begin{align}
    \int_{\mathbb{T}^M} f_{-}^{(\epsilon)}(\bx) d\bx \leq \liminf_{M \to \infty}  \frac{1}{M} \sum_{n=1}^{M}\chi_{B(\by, r)}(n \ba) \leq 
    \limsup_{M \to \infty}  \frac{1}{M} \sum_{n=1}^{M}\chi_{B(\by, r)}(n \ba) \leq \int_{\mathbb{T}^M} f_{+}^{(\epsilon)}(\bx) d\bx.
  \end{align}
  We take $\epsilon \to 0$; then $\int_{\mathbb{T}^M} f_{\pm}^{(\epsilon)}(\bx) d\bx \to \int_{\mathbb{T}^M} \chi_{B(\by, r)}(\bx) d\bx$ and
  \begin{align}
    \liminf_{M \to \infty}  \frac{1}{M} \sum_{n=1}^{M}\chi_{B(\by, r)}(n \ba) = \limsup_{M \to \infty}  \frac{1}{M} \sum_{n=1}^{M}\chi_{B(\by, r)}(n \ba) = \int_{\mathbb{T}^M} \chi_{B(\by, r)}(\bx) d\bx.
  \end{align}
  Therefore, $ \lim_{M \to \infty}  \frac{1}{M} \sum_{n=1}^{M}\chi_{B(\by, r)}(n \ba) = \int_{\mathbb{T}^M} \chi_{B(\by, r)}(\bx) d\bx$. 
  Because $\int_{\mathbb{T}^M} \chi_{B(\by, r)}(\bx) d\bx > 0$ with $r>0$, then $\lim_{M \to \infty}  \frac{1}{M} \sum_{n=1}^{M}\chi_{B(\by, r)}(n \ba) > 0$. Therefore, with an arbitrary $\by\in\bbT^M$ and an arbitrary $r>0$, there exists $n \in \bbN$ such that $\chi_{B(\by, r)}(n \ba) > 0$ or $n \ba \in B(\by, r)$. 
  This result proves that $\{n\ba \in \mathbb{T}^M \mid n \in \mathbb{N}\}$ is dense in $\mathbb{T}^M$.

\section{Evaluation of the approximation rate}
An interesting theoretical question is how to describe relative goodness or badness in a universal approximation.
The approximation rate can be used here, which is the decay rate of the approximation error.
This rate refers to the speed at which the approximation error decreases when the parameters, such as the number $N$ of qubits and the input dimension $d$, are increased.
The approximation rate strongly depends on the nature of the target function $g$ to be approximated.
Here, we provide the evaluation of the approximation rate in the parallel scenario.
Let us mention the circuit  $
    \cV_N(\bx) = V_1(\bx) \otimes V_2(\bx) \otimes \ldots \otimes V_N(\bx),
    $
where
$
    V_j(\bx) = e^{-i \textup{arccos}(\sqrt{x_{[j]}})Y}, (1\leq j \leq N)
$
with $[j]$ denoting the integer number $k$ such that $1\leq k \leq d$ and $j \equiv k (\textup{mod } d)$.
For the sake of simplicity, we consider the case when $N$ is a multiple of $d$ as $N=\np d$.

For each combination of $(p_1,p_2,\ldots,p_d)$, where $p_i=0,\ldots,\np$, we show that we can create the term $\prod_{i=1}^d x_i^{p_i}(1-x_i)^{\np-p_i}$ in the basis functions. Indeed, for each $j=1,\ldots,N$, we have
\begin{align}
        \bra{0} {V_{j}}^{\dagger}(\bx) (\dfrac{I + Z}{2}) V_j(\bx)\ket{0} &= \left(\bra{0} {V_{j}}^{\dagger}(\bx) I V_j(\bx)\ket{0} + \bra{0} {V_{j}}^{\dagger}(\bx) Z V_j(\bx)\ket{0}\right)/2 \\
        &= \left(1 + (2x_{[j]}-1)\right)/2 = x_{[j]}.
    \end{align}
Similarly, we also have 
    \begin{align}
        \bra{0} {V_{j}}^{\dagger}(\bx) (\dfrac{I - Z}{2}) V_j(\bx)\ket{0} &= \left(\bra{0} {V_{j}}^{\dagger}(\bx) I V_j(\bx)\ket{0} - \bra{0} {V_{j}}^{\dagger}(\bx) Z V_j(\bx)\ket{0}\right)/2 \\
        &= \left(1 - (2x_{[j]}-1)\right)/2 = 1 - x_{[j]}.
    \end{align}
For all $j=1,\ldots,nd$, we can write $j=i + kd$, where $i=1,\ldots,d$ and $k=0,\ldots,\np-1$.
If we denote the following operators
\begin{align}
    A_j = A_{i+kd} = \begin{cases} (I + Z)/2 & \textup{ if } k = 0,\ldots, p_i-1 
    \\  (I - Z)/2 & \textup{ if } k = p_i,\ldots,\np-1 \end{cases} 
\end{align}
then,
\begin{align}
  \bra{0} {V_{j}}^{\dagger}(\bx) A_j V_j(\bx)\ket{0}  = \begin{cases} x_{i} & \textup{ if } k = 0,\ldots, p_i-1
    \\  1-x_{i} & \textup{ if } k = p_i,\ldots,\np-1. \end{cases} 
\end{align}

Therefore, if we consider the observable $O_{\bp}=A_1\otimes A_2\otimes \ldots \otimes A_N$ (where $\bp$ denotes $(p_1,p_2,\ldots,p_d)$), then the basis function $\psi_{\bp}$ corresponding with $O_{\bp}$ becomes
\begin{align}
    \psi_{\bp} = \bra{0}^{\otimes N} \cV_N^{\dagger}(\bx) O_{\bp} \cV_N(\bx) \ket{0}^{\otimes N} &= \bra{0}^{\otimes N} \left(V_1(\bx) \otimes \ldots \otimes V_N(\bx)\right)^{\dagger} O_{\bp} \left(V_1(\bx) \otimes \ldots \otimes V_N(\bx)\right) \ket{0}^{\otimes N}\\
    &= \prod_{j=1}^{N} \bra{0} {V_{j}}^{\dagger}(\bx) A_j V_j(\bx)\ket{0}\\
    &= \prod_{i=1}^d x_i^{p_i}(1-x_i)^{\np-p_i}\label{eqn:basis:bernstein}.
\end{align}

Based on this selection of observables, we show that we can obtain UAP for our quantum feature framework in the parallel scenario.
Now, consider a continuous function $g$ on the compact set $\cX \subset [0, 1]^d$. 
Let $\bp=(p_1, \ldots, p_d)$, and then $\dfrac{\bp}{\np}\in[0,1]^d$.
We consider the output weight $w_{\bp}=g(\frac{\bp}{\np})\prod_{i=1}^d \binom{\np}{p_i}$.
The output function becomes the multivariate Bernstein polynomial
\begin{align}
    f(\bx)=P(\bx) = \sum_{\bp}w_{\bp}\psi_{\bp} = \sum_{p_1=0}^{\np}\ldots\sum_{p_d=0}^{\np} g(\frac{\bp}{\np})\prod_{i=1}^d \binom{\np}{p_i}x_i^{p_i}(1-x_i)^{\np-p_i}.
\end{align}
The approximation of multivariate Bernstein polynomials to a continuous function has been explored before, for example, in Refs~\cite{heitz:2002:multivariate,wouodjie:2014:multivariate,foupouagnigni:2020:multivariate}. Here, we provide the proof related to this approximation property, then present our own result in the evaluation of approximation rate.

Because $\cX$ is a compact set and $g$ is uniformly continuous with respect to the Euclidean norm $\Vert \cdot \Vert_2$,
$g$ admits a modulus of continuity $\omega : [0, \infty) \to [0, \infty)$ such that 
\begin{align}
\omega(\delta)=\sup_{\Vert \bx -\by \Vert_{2} \leq \delta}|g(\bx)-g(\by)| \text{ for $\delta \geq 0$,}
\end{align}
where $\lim_{\delta \to 0}\omega(\delta)=0$.

We propose and prove the following result, which shows the UAP of our quantum feature framework in this setting.

\begin{result}[UAP in the parallel scenario with the sufficient number of qubits]\label{theo:eval:qubits}
For any continuous function $g: \cX \to \bbR$ and the above-mentioned construction of output function $P(\bx)$, then for any $\varepsilon > 0$ there exist $N=\np d$ such that
$
    |g(\bx) - P(\bx)| < \varepsilon
$
for all $\bx \in \cX$.
Let $M$ be an arbitrary positive number such that $M\geq\omega(\sqrt{d})$, then all $\np > m(\varepsilon)$ satisfy our statement, where 
\begin{align}
m(\varepsilon) = \inf_{\delta > 0, \omega(\delta) < \varepsilon}\left[ \dfrac{Md^2}{2(\varepsilon - \omega(\delta))\delta^2} \right].
\end{align}
Furthermore, if the target function $g$ is Lipschitz continuous with respect to the Euclidean norm, we can take the number of qubits $N=dn=O(d^{7/2}\varepsilon^{-3})$ to obtain an $\varepsilon$-approximator for $g$.
Equivalently, we can write $\varepsilon = O(d^{7/6}N^{-1/3})$, which describes the approximation rate in our quantum feature framework.
\end{result}

\paragraph{Proof.---}
First, we prove the following lemmas.
\begin{lemma}[\cite{yoshida:1980:functional}~pp.~8--9]\label{lemma:binom}
    For $x \in [0, 1]$, consider $r_p(x) = \binom{n}{p}x^p (1-x)^{n-p}$; then 
    \begin{equation}\label{eqn:form}
        \sum_{p=0}^n (p-nx)^2 r_p(x) = nx(1-x).
    \end{equation}
\end{lemma}
\paragraph{Proof for lemma~\ref{lemma:binom}.---}To prove this lemma, we consider the following formula:
\begin{align}\label{eqn:binom}
    (x+y)^n = \sum_{p=0}^n \binom{n}{p}x^p y^{n-p}.
\end{align}
We take the zero, one and second-order derivative of two sides of Eq.~\eqref{eqn:binom} by $x$ and let $y=1-x$ in each equation.
Then we have $\sum_{p=0}^n r_p(x) = 1, \sum_{p=0}^n pr_p(x) = nx$ and $\sum_{p=0}^n p(p-1)r_p(x) = n(n-1)x^2$.
Therefore, we can obtain Eq.~\eqref{eqn:form} as follows:
\begin{align}
    \sum_{p=0}^n (p-nx)^2 r_p(x) &= \sum_{p=0}^n p(p-1)r_p(x) - (2n-1)\sum_{p=0}^n pr_p(x) + n^2x^2\sum_{p=0}^nr_p(x)\\
    &= n(n-1)x^2 - (2n-1)nx + n^2x^2 = nx(1-x).
\end{align}

\begin{lemma}[\cite{davis:1975:interpolation}~pp.~110--111]\label{lemma:ineq}
    For a given $\delta > 0$ and $x \in [0, 1]$ we have
    \begin{equation}\label{eqn:ineq}
        \sum_{\left|\frac{p}{n}-x\right| \geq \delta} \binom{n}{p}x^p (1-x)^{n-p} \leq \dfrac{1}{4n\delta^2},
    \end{equation}
    where the sum is taken over those values of $p=0,1,\ldots,n$ for which $\left| \dfrac{p}{n} - x\right| \geq \delta$.
\end{lemma}
\paragraph{Proof for lemma~\ref{lemma:ineq}.---}
If $\left|\dfrac{p}{n}-x\right| \geq \delta$ then $\left( \dfrac{p-nx}{n\delta} \right)^2 \geq 1$. Hence, we have
\begin{align}
    \sum_{\left|\frac{p}{n}-x\right| \geq \delta} \binom{n}{p}x^p (1-x)^{n-p} &\leq \sum_{\left|\frac{p}{n}-x\right| \geq \delta} \left( \dfrac{p-nx}{n\delta} \right)^2\binom{n}{p}x^p (1-x)^{n-p}\\
    &\leq \sum_{p=0}^n \left( \dfrac{p-nx}{n\delta} \right)^2\binom{n}{p}x^p (1-x)^{n-p}\\
    &= \left( \dfrac{1}{n\delta}\right)^2\sum_{p=0}^n (p-nx)^2r_p(x)\\
    &= \left( \dfrac{1}{n\delta}\right)^2 nx(1-x) \textup{ \quad (use lemma~\ref{lemma:binom})}\\
    &\leq \dfrac{1}{4n\delta^2} \textup{ \quad (since $x(1-x)\leq \dfrac{1}{4}$ for $x\in [0,1]$)}.
\end{align}

Next, we prove Result~\ref{theo:eval:qubits}. Consider $\delta > 0$; then we have 
    \begin{align}
        \left| g(\bx) - P(\bx)  \right| &= \left| g(\bx) - \sum_{p_1=0}^{\np} \cdots \sum_{p_d=0}^{\np} g\left(\frac{\bp}{\np}\right) \prod_{i=1}^{d}\binom{\np}{p_i}x_i^{p_i} (1-x_i)^{\np-p_i} \right| \\
        &= \left| \sum_{p_1=0}^{\np} \cdots \sum_{p_d=0}^{\np} \left(g(\bx) - g\left(\frac{\bp}{\np}\right)\right) \prod_{i=1}^{d}\binom{\np}{p_i}x_i^{p_i} (1-x_i)^{\np-p_i} \right| \\
        &\leq \left| \sum_{\bp, ||\frac{\bp}{\np} - \bx||_{\infty} < \frac{\delta}{\sqrt{d}}} \mbox{〃} \right|  + \left| \sum_{\bp, ||\frac{\bp}{\np} - \bx||_{\infty} \geq \frac{\delta}{\sqrt{d}}} \mbox{〃} \right|\\
        &\leq \underbrace{\left| \sum_{\bp, ||\frac{\bp}{\np} - \bx||_{2} \leq \delta} \mbox{〃} \right|}_{E_1}  + \underbrace{\left| \sum_{\bp, ||\frac{\bp}{\np} - \bx||_{\infty} \geq \frac{\delta}{\sqrt{d}}} \mbox{〃} \right|}_{E_2}\label{eqn:two:sides}. 
    \end{align}
    
    We evaluate the first term $E_1$ on the right side of Eq.~\eqref{eqn:two:sides} using the uniformly continuous property of $g$ 
    \begin{align}
        E_1 &\leq \omega(\delta) \sum_{\bp, ||\frac{\bp}{\np} - \bx||_{2} \leq \delta} \prod_{i=1}^{d}\binom{\np}{p_i}x_i^{p_i} (1-x_i)^{\np - p_i}\\
        &\leq \omega(\delta) \sum_{p_1=0}^{\np} \cdots \sum_{p_d=0}^{\np} \prod_{i=1}^{d}\binom{\np}{p_i}x_i^{p_i} (1-x_i)^{\np - p_i}\\
        & = \omega(\delta).
    \end{align}
    
    Next, we evaluate $E_2$ as follows:
    \begin{align}
        E_2 &\leq \sum_{\bp, ||\frac{\bp}{\np} - \bx||_{\infty} \geq \frac{\delta}{\sqrt{d}}} \left| g(\bx) - g\left(\frac{\bp}{\np}\right) \right|  \prod_{i=1}^{d}\binom{\np}{p_i}x_i^{p_i} (1-x_i)^{\np - p_i}\\
        &\leq \omega(\sqrt{d}) \sum_{\bp, ||\frac{\bp}{\np} - \bx||_{\infty} \geq \frac{\delta}{\sqrt{d}}}  \prod_{i=1}^{d}\binom{\np}{p_i}x_i^{p_i} (1-x_i)^{\np - p_i}\label{eqn:eval:mod}\\
        &\leq M \sum_{k=1}^{\np}\sum_{\bp \in \Omega_k}  \prod_{i=1}^{d}\binom{\np}{p_i}x_i^{p_i} (1-x_i)^{\np - p_i},\label{eqn:e2:last}
    \end{align}
    where $\Omega_k$ is the collection of $\bp$ such that there are $k$ number $p_i$ in $p_1,\ldots,p_d$ that satisfy $|\dfrac{p_i}{\np} - x_i| \geq \frac{\delta}{\sqrt{d}}$.
    Equation~\eqref{eqn:eval:mod} comes from the fact that $\left| g(\bx) - g\left(\frac{\bp}{\np}\right) \right| \leq \omega(\sqrt{d})$
    since $||\bx - \frac{\bp}{\np}||_{2} \leq \sqrt{d}$.
    
    We note that
    \begin{align}
        \sum_{\left| \frac{p_i}{\np} - x_i\right| < \frac{\delta}{\sqrt{d}}}\binom{\np}{p_i}x_i^{p_i} (1-x_i)^{\np - p_i} \leq \sum_{p_i=0}^{\np}\binom{\np}{p_i}x_i^{p_i} (1-x_i)^{\np - p_i} = 1,
    \end{align}
    and from lemma~\ref{lemma:ineq}
        \begin{align}
        \sum_{\left| \frac{p_i}{\np} - x_i\right| \geq \frac{\delta}{\sqrt{d}}}\binom{\np}{p_i}x_i^{p_i} (1-x_i)^{\np - p_i} \leq \dfrac{d}{4\np \delta^2}.
    \end{align}
    Therefore, from Eq.~\eqref{eqn:e2:last} we have
    \begin{align}
        E_2 \leq M\sum_{k=1}^{d}\binom{d}{k}\left( \dfrac{d}{4\np \delta^2} \right)^k
        = M \left( \left( 1 + \dfrac{d}{4\np \delta^2}\right)^{d} - 1\right).
    \end{align}
    Therefore, 
    \begin{equation}
        \left| g(\bx) - P(\bx)  \right| \leq E_1 + E_2 \leq \omega(\delta) + M \left( \left( 1 + \dfrac{d}{4\np \delta^2}\right)^{d} - 1\right).
    \end{equation}
    
    To obtain the UAP, since $\lim_{\delta \to 0^{+}}\omega(\delta)=0$, we select $n$ and $\delta$ such that $\omega(\delta) < \varepsilon$ and
    \begin{align}
         \omega(\delta) + M \left( \left( 1 + \dfrac{d}{4\np \delta^2}\right)^{d} - 1\right) < \varepsilon,
    \end{align}
    or it is equivalent to
    \begin{align}\label{eqn:n:condition}
        n > \dfrac{d}{4\delta^2} \dfrac{1}{\left( 1 + \dfrac{\varepsilon - \omega(\delta)}{M}\right)^{1/d} - 1}.
    \end{align}
    Because $e^u > 1 + u$ and $\ln(1+u) > \dfrac{2u}{2+u}$ for all $u > 0$, let $\eta=\omega(\delta)$, we have
    \begin{align}
        \left(1 + \dfrac{\varepsilon - \eta}{M}\right)^{1/d} &= e^{\ln\left(1+\dfrac{\varepsilon - \eta}{M}\right)/d} \\
        &> 1 + \dfrac{1}{d} \ln\left(1+\dfrac{\varepsilon - \eta}{M}\right) \textup{\quad (use $e^u > 1 + u$)}\\
        &> 1 + \dfrac{1}{d}\dfrac{2\dfrac{\varepsilon-\eta}{M}}{2 + \dfrac{\varepsilon-\eta}{M}} \textup{\quad (use $\ln(1+u) > \dfrac{2u}{2+u}$)} \\
        &= 1 + \dfrac{1}{d}\dfrac{2(\varepsilon - \eta)}{2M + (\varepsilon - \eta)}\\
        &> 1 + \dfrac{\varepsilon - \eta}{2Md},
    \end{align}
    where we choose $\varepsilon$ such that $\varepsilon < 2M$.
    Hence,
    \begin{align}
    \dfrac{d}{4\delta^2} \dfrac{1}{\left( 1 + \dfrac{\varepsilon - \omega(\delta)}{M}\right)^{1/d} - 1} <  \dfrac{2Md^2}{4\left(\varepsilon - \omega(\delta)\right)\delta^2} = \dfrac{Md^2}{2\left(\varepsilon - \omega(\delta)\right)\delta^2} = s(\delta).
\end{align}
If we choose $n > m(\varepsilon) = \inf_{\delta > 0, \omega(\delta) < \varepsilon} s(\delta)$, there always exists $\delta > 0$ that satisfies $\omega(\delta) < \varepsilon$ and Eq.~\eqref{eqn:n:condition}.
Therefore, the first part of Result~\ref{theo:eval:qubits} is proven.

Next, we prove the second part of Result~\ref{theo:eval:qubits}. If $g$ is Lipschitz continuous with respect to the Euclidean norm, there exists some constant $C > 0$ such that $\omega(\delta)\leq C\delta$.
For $0 < \delta < \varepsilon / C$, then $\varepsilon - \omega(\delta) \geq \varepsilon - C\delta > 0$. Since $C\sqrt{d}\geq \omega(\sqrt{d})$, we can choose $M=C\sqrt{d}$, then
\begin{align}
    s(\delta) = \dfrac{Md^2}{2\left(\varepsilon - \omega(\delta)\right)\delta^2} \leq \dfrac{Cd^{5/2}}{2 (\varepsilon - C\delta)\delta^2}.
\end{align}

Therefore, to obtain the UAP, we only need to choose 
\begin{align}
n > \inf_{0<\delta  < \varepsilon/C} \left[\dfrac{Cd^{5/2}}{2 (\varepsilon - C\delta)\delta^2}\right].
\end{align}

We note that
\begin{align}
    \dfrac{Cd^{5/2}}{2 (\varepsilon - C\delta)\delta^2} &= \dfrac{d^{5/2}}{8}\dfrac{1}{(\frac{\varepsilon}{C} - \delta)(\frac{\delta}{2})(\frac{\delta}{2})} \\
    &\geq \dfrac{d^{5/2}}{8}\dfrac{27}{(\frac{\varepsilon}{C} - \delta + \frac{\delta}{2} +  \frac{\delta}{2})^3} \textup{\quad (use $\dfrac{1}{xyz} \geq \left( \dfrac{3}{x+y+z}\right)^3$ for all $x, y, z > 0$)}\\
    &= \dfrac{27C^3d^{5/2}}{8\varepsilon^3}.
\end{align}

The equality occurs when $\delta = \bar{\delta}=\dfrac{2\varepsilon}{3C}$ that satisfies $\bar{\delta} < \dfrac{\varepsilon}{C}$. Then we have 
\begin{align}
\inf_{0<\delta  < \varepsilon/C} \left[\dfrac{Cd^{5/2}}{2 (\varepsilon - C\delta)\delta^2}\right]=\dfrac{27C^3d^{5/2}}{8\varepsilon^3}.
\end{align}

Therefore, we can take the number of qubits $N=dn=O(d^{7/2}\varepsilon^{-3})$ to obtain an $\varepsilon$-approximator for the target function $g$ if $g$ is Lipschitz continuous.
Furthermore, we can write $\varepsilon = O(d^{7/6}N^{-1/3})$, which describes the approximation rate in our quantum feature framework. 

Finally, we demonstrate in the following result that we can have a better approximation rate and a better evaluation for the number of qubits to obtain an $\varepsilon$-approximator.

\begin{result}[Better approximation rate]\label{theo:better:qubits}
For any continuous function $g: \cX \to \bbR$, there exist an output function $f(\bx)$ of our quantum feature framework with $N$ qubits and some constant $A$ such that
\begin{align}
    |g(\bx) - f(\bx)| < A\omega(\dfrac{d^{3/2}}{N})
\end{align}
for all $\bx \in \cX$.
Then for any $\varepsilon > 0$, we can choose the number of qubits $N$ such that $A\omega(\dfrac{d^{3/2}}{N}) < \varepsilon$ to obtain an $\varepsilon$-approximator for $g(\bx)$.
Furthermore, if $g$ is Lipschitz continuous with Lipschitz constant $C$ with respect to the Euclidean norm, we can take $N > ACd^{3/2}\varepsilon^{-1}$.
In this case, the approximation error becomes $\varepsilon=O(d^{3/2}N^{-1})$.
\end{result}

\paragraph{Proof.---}
We use the Jackson theorem~\cite{newman:1964:jackson} in higher dimension of the quantitative information on the degree of polynomial approximation to a continuous function.
Let $\cP_{N}^d$ denote the linear space of $d$-variate polynomials of degree less than or equal to $N$.
Theorem 4 in Ref.~\cite{newman:1964:jackson} states that there exists a constant $A$ and a multivariate polynomial $P(\bx)\in \cP_{N}^d$ such that 
\begin{align}
    |g(\bx) - P(\bx)| < A\omega(\dfrac{d^{3/2}}{N})
\end{align}
for all $\bx \in \cX$.
As the Bernstein basis polynomials in Eq.~\eqref{eqn:basis:bernstein} form a basis of $\cP_{N}^d$,
there always exists an output function $f(\bx)$ in our quantum feature framework, which is the linear combination of the Bernstein basis polynomials in Eq.~\eqref{eqn:basis:bernstein} that satisfies $f(\bx)=P(\bx)$.

If $g$ is Lipschitz continuous with Lipschitz constant $C$, then $A\omega(\dfrac{d^{3/2}}{N})\leq \dfrac{ACd^{3/2}}{N}$.
Hence, if we choose $N > ACd^{3/2}\varepsilon^{-1}$, then $|g(\bx) - f(\bx)| < \dfrac{ACd^{3/2}}{N} < \varepsilon$,
and $f(\bx)$ becomes an $\varepsilon$-approximator of $g(\bx)$.
In this case, we can take the number of qubits $N=O(d^{3/2}\varepsilon^{-1})$ to obtain an $\varepsilon$-approximator for the target function $g$. The approximation error becomes $\varepsilon = O(d^{3/2}N^{-1})$.

We obtain Result 4 in the main text from Result~\ref{theo:eval:qubits} and Result~\ref{theo:better:qubits}. 
Result~\ref{theo:better:qubits} provides the better approximation rate than Result~\ref{theo:eval:qubits} in terms of the number of qubits $N$, because if we increase $N$, the approximation error will reduce faster.
However, Result~\ref{theo:eval:qubits} is better in terms of the input dimension $d$, because if we increase $d$, the approximation error in Result~\ref{theo:eval:qubits} will increase slower.
We note that Result~\ref{theo:better:qubits} does not provide the explicit form of the approximator as in Result~\ref{theo:eval:qubits}.
The explicit form of the approximator in Result~\ref{theo:better:qubits} remains for future work.

\providecommand{\noopsort}[1]{}\providecommand{\singleletter}[1]{#1}%